\documentclass[conference]{IEEEtran}
\usepackage{cite}
\usepackage[pdftex]{graphicx}
\DeclareGraphicsExtensions{.pdf,.jpeg,.png,.JPG,.jpg} 
\usepackage{amsmath}
\usepackage[caption=false,font=footnotesize]{subfig} 
\usepackage{url}
\usepackage{fancyhdr}

\begin{document}

\title{Anomalous Frequency Noise from the Megahertz Channelizing Resonators in Frequency-Division Multiplexed Transition Edge Sensor Readout}

\author{\IEEEauthorblockN{John Groh\IEEEauthorrefmark{1},
Kam Arnold\IEEEauthorrefmark{2},
Jessica Avva\IEEEauthorrefmark{1},
Darcy Barron\IEEEauthorrefmark{3},
Kevin T. Crowley\IEEEauthorrefmark{1},
Matt Dobbs\IEEEauthorrefmark{4}\IEEEauthorrefmark{5},
Tijmen de Haan\IEEEauthorrefmark{6},\\
William Holzapfel\IEEEauthorrefmark{1},
Adrian Lee\IEEEauthorrefmark{1}\IEEEauthorrefmark{7},
Lindsay Ng Lowry\IEEEauthorrefmark{2},
Joshua Montgomery\IEEEauthorrefmark{4},
Maximiliano Silva-Feaver\IEEEauthorrefmark{2},\\
Aritoki Suzuki\IEEEauthorrefmark{7}, and
Nathan Whitehorn\IEEEauthorrefmark{8}}
  \IEEEauthorblockA{\IEEEauthorrefmark{1}Department of Physics, University of California, Berkeley, CA 94720, USA}
  \IEEEauthorblockA{\IEEEauthorrefmark{2}Department of Physics, University of California, San Diego, La Jolla, CA 92093-0424, USA}
  \IEEEauthorblockA{\IEEEauthorrefmark{3}Department of Physics and Astronomy, University of New Mexico, Albuquerque, NM 87131, USA}
  \IEEEauthorblockA{\IEEEauthorrefmark{4}Physics Department, McGill University, Montreal, QC H3A 0G4, Canada}
  \IEEEauthorblockA{\IEEEauthorrefmark{5}Canadian Institute for Advanced Research (CIFAR), Toronto, ON M5G 1M1, Canada}
  \IEEEauthorblockA{\IEEEauthorrefmark{6}High Energy Accelerator Research Organization (KEK), Tsukuba, Ibaraki 305-0801, Japan}
  \IEEEauthorblockA{\IEEEauthorrefmark{7}Physics Division, Lawrence Berkeley National Laboratory, Berkeley, CA 94720, USA}
  \IEEEauthorblockA{\IEEEauthorrefmark{8}Department of Physics and Astronomy, University of California, Los Angeles, CA 90095, USA}
}

\maketitle

\begin{abstract}
Superconducting lithographed resonators have a broad range of current and potential applications in the multiplexed readout of cryogenic detectors.  Here, we focus on LC bandpass filters with resonances in the 1-5 MHz range used in the transition edge sensor (TES) bolometer readout of the Simons Array cosmic microwave background (CMB) experiment.  In this readout scheme, each detector signal amplitude-modulates a sinusoidal carrier tone at the resonance frequency of the detector's accompanying LC filter.  Many modulated signals are transmitted over the same wire pair, and quadrature demodulation recovers the complex detector signal.  We observe a noise in the resonant frequencies of the LC filters, which presents primarily as a current-dependent noise in the quadrature component after demodulation.  This noise has a rich phenomenology, bearing many similarities to that of two-level system (TLS) noise observed in similar resonators in the GHz regime.  These similarities suggest a common physical origin, thereby offering a new regime in which the underlying physics might be probed.  We further describe an observed non-orthogonality between this noise and the detector responsivities, and present laboratory measurements that bound the resulting sensitivity penalty expected in the Simons Array.  From these results, we do not anticipate this noise to appreciably affect the overall Simons Array sensitivity, nor do we expect it to limit future implementations.
\end{abstract}


%
\IEEEpeerreviewmaketitle

\fancyhf{} 
\renewcommand{\headrulewidth}{0pt}
\thispagestyle{fancy}
\pagestyle{empty}
\lfoot{{\small \copyright 2021 IEEE.  Personal use of this material is permitted.  Permission from IEEE must be obtained for all other uses, in any current or future media, including reprinting/republishing this material for advertising or promotional purposes, creating new collective works, for resale or redistribution to servers or lists, or reuse of any copyrighted component of this work in other works.}}

\section{Introduction}

\par
Multiplexed readout is an increasingly common requirement for modern cryogenic detector arrays, as it reduces cost, wiring complexity, and cooling requirements.  A natural direction toward multiplexing is with superconducting lithographed resonators, which provide many cryogenically scalable paths to multiplex signals in frequency space.  Many such demonstrations exist in the literature in varying forms of maturity: with amplitude readout of resonances modulated by a changing detector impedance \cite{Dobbs2012}, with kinetic inductance detectors which double as readout resonators \cite{MKID}, with inductively coupled RF SQUIDs that modulate the resonance frequencies \cite{uMux}, and with a high-kinetic inductance resonator to transduce low-frequency currents into RF resonance frequency shifts \cite{MKING}.  Together, these frequency multiplexing techniques have been used to read out a variety of detectors, including TES calorimeters and bolometers, metallic magnetic calorimeters, superconducting nanowire single photon detectors, and many flavors of kinetic inductance detectors \cite{Dobbs2012, TESCalorimeterReview, MMC_uMux, MKING, ZmuidzinasReview}.

\par
\begin{figure}[!t]
  \centering
  \includegraphics[width=\linewidth]{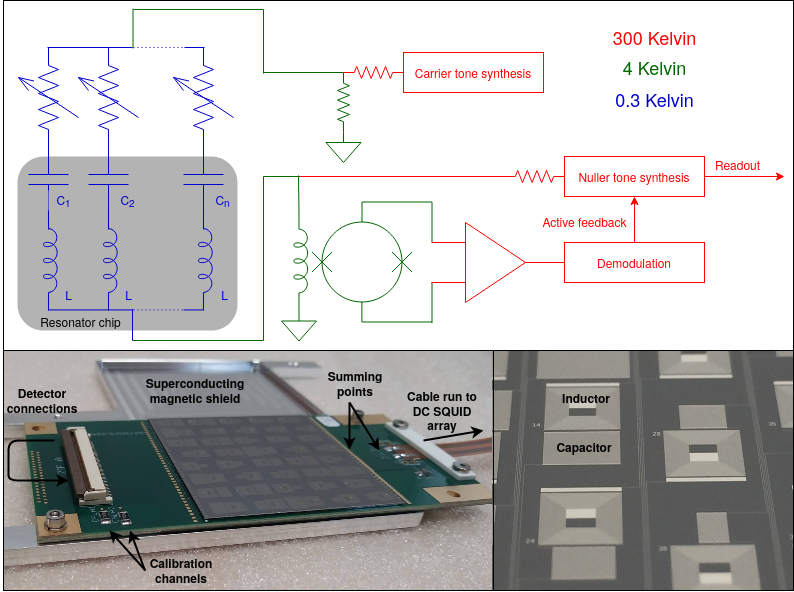}
  \caption{\textit{Top:} Simplified schematic of the DfMux readout system.  \textit{Bottom left:} A Simons Array resonator chip along with its mounting board and magnetic shielding.  The cable on the right transmits the summed signals from 40 channels to the SQUID array for amplification.  The connector on the left provides interconnects between the resonators and TES bolometers.  Visible on the lower left of the board are two 1 $\Omega$ calibration resistors, which were used for many measurements presented here.  \textit{Bottom right:} A close view of the individual resonator elements.  As the spiral inductors require a wire bond for completion, they are identifiable by their central and exterior bond pads.  The interdigitated capacitors are seen to vary in size, as the resonanance frequencies are specified by varying the channel capacitances.}
  \label{fig_LCs}
\end{figure}

In this manuscript, we describe a phenomena observed in an implementation of the first aforementioned multiplexing paradigm for TES bolometer readout; in particular, the Digital Frequency-domain Multiplexing (DfMux) readout system for the Simons Array CMB polarization experiment \cite{SimonsArray}.  In this readout system, depicted schematically at the top of Fig. \ref{fig_LCs}, each detector is connected in series with an LC bandpass filter so that it can be uniquely addressed by a carrier tone, which doubles as the detector bias and becomes amplitude modulated by the detector signal.  A DC Superconducting QUantum Interference Device (SQUID) array, well-matched to the TES impedances, senses the time-varying current produced by the voltage biased detectors.  Active nulling feedback acts to linearize the SQUIDs and greatly increases the dynamic range of the system \cite{DAN_DEVB}.  Custom digital electronics provide carrier synthesis and demodulation at ambient temperature \cite{ICE}.

\par
The Simons Array detector readout system was built on the heritage of previous implementations in the POLARBEAR \cite{PB1} and SPTpol \cite{SPTpol} CMB experiments, and was jointly developed for the SPT-3G \cite{SPT3G} experiment.  Several technology advances and architectural changes were crucial for scaling up the previously achieved multiplexing factors of 8x (16x) in POLARBEAR (SPTpol) to 40x (68x) in the Simons Array (SPT-3G), a few of which are relevant for the noise mechanism discussed in this manuscript.  To accommodate more channels per multiplexer, the channelizing resonators were designed with roughly 4.5x higher resonance frequencies and 4.5x narrower bandwidths.  For scalability and greater control over resonance frequency placement, the construction of these LC resonators was migrated to a single lithographically defined monolithic chip per multiplexer \cite{KajaLCProceedings}.  The resonators take the form of lumped-element single-layer interdigitated capacitors and spiral inductors completed by a wire bond as shown in the bottom panels of Fig. \ref{fig_LCs}, and are fabricated with a superconducting film of either aluminum or niobium.  Under observation conditions, the total quality factors are $\mathcal{O}(10^3)$.  All data presented hereafter were collected using Simons Array readout hardware.\footnote{While the Simons Array and SPT-3G readout systems are identical in concept, minor implementation differences are important for the level of the noise mechanism discussed in this manuscript; any results here should not be assumed to represent phenomena in the SPT-3G system at a quantitative level.}

\section{Resonator Frequency Noise}

\par
A key requirement on any readout system for CMB detectors is that its noise contribution be sub-dominant to the noise from statistical fluctuations of the detected photons.  When referenced to a current at the SQUID array input coil, we expect a photon white noise level of roughly 30 pA/$\sqrt{\mathrm{Hz}}$ in the Simons Array, and we target a readout noise level which does not appreciably increase the total noise level.

\par
Due to the quadrature demodulation in use, we expect different noise levels in the in-phase ($I$) and quadrature ($Q$) timestreams.  The demodulation electronics choose the phase that defines the ($I$,$Q$) basis after detector tuning by enforcing that $\langle Q(t)\rangle = 0$ so that the carrier and nuller signals are entirely in $I$.  In this basis, $I$ and $Q$ have a simple interpretation as the real and imaginary parts of the detector current.  Noise sources that are power noise sources in origin will then appear preferentially in $I$, as they are intrinsically aligned with the bolometer responsivities.  Reactive parasitics in the cryogenic circuitry are expected to cause the detector signals and power noise sources to be rotated from this default ($I$,$Q$) basis by a few degrees.  After applying a small offline rotation to a new ($I$',$Q$') basis that undoes this effect, we expect to use the $I$' time-ordered data for science analysis and discard $Q$'.  Noise sources not dependent on the detector responsivity (e.g. SQUID noise, amplifier noise, resistor Johnson noise) were expected at the time of design to be equal in amplitude between $I$ and $Q$.  A minor exception is the TES Johnson noise, which is preferentially suppressed by the detector loop gain in $I$ but only contributes a negligible fraction of the total noise budget.

\par
\begin{figure}[!t]
  \centering
  \subfloat{\includegraphics[width=0.62\linewidth]{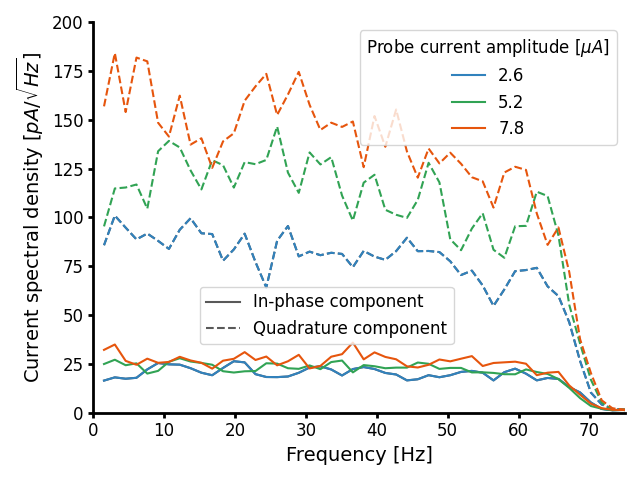}}
  \subfloat{\includegraphics[width=0.38\linewidth]{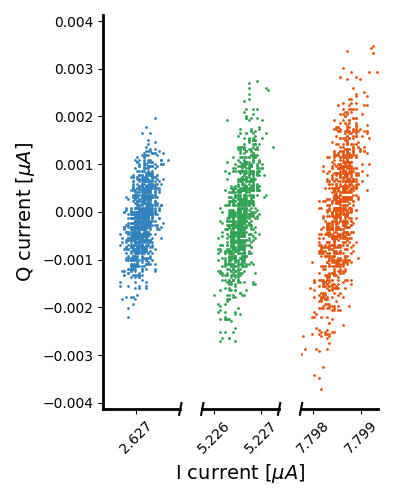}}
  \caption{\textit{Left:} Spectral densities of the $I$ and $Q$ components of the current through a 3.4 MHz, 1.3 $\Omega$ channel at $T$ = 0.6 K at several probe currents.  The roll off around 70 Hz is due to an anti-aliasing filter in the demodulation electronics.  \textit{Right:} Data from three measurements shown on the left, plotted in the I-Q plane.  The relatively small probe current dependence in the $I$ noise is evident as simply a consequence of the complex angle of the asymmetric noise, and is removable by a change of basis.  The relatively high temperature of 0.6 K enables the use of normal-state TESs for readout characterization, an outlier of which was selected here for visual clarity, but the noise phenomena are qualitatively unchanged at the nominal operating temperature of 0.3 K.}
  \label{fig_noise_phenomenology}
\end{figure}

However, instead of an equivalency between readout-induced noise in $I$ and $Q$, we observe a large excess of $Q$ noise.  Additionally, this excess increases with the current through the resonator and is coupled with a much smaller current dependence in the $I$ noise level.  These effects are demonstrated in Fig. \ref{fig_noise_phenomenology}, in which a resistive channel with zero power-to-current responsivity was ``biased'' at several current amplitudes.  As can be seen in the right part of the figure, there is a small correlation between $I$ and $Q$, explaining the current dependence of the $I$ noise via a phase shift between the carrier as measured by the room temperature electronics and the phase at which this excess noise is injected.

\par
\begin{figure}[!t]
  \centering
  \subfloat{\includegraphics[width=0.5\linewidth]{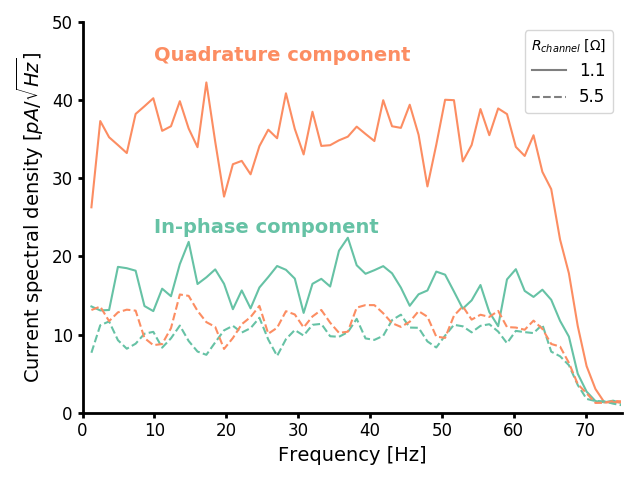}}
  \subfloat{\includegraphics[width=0.5\linewidth]{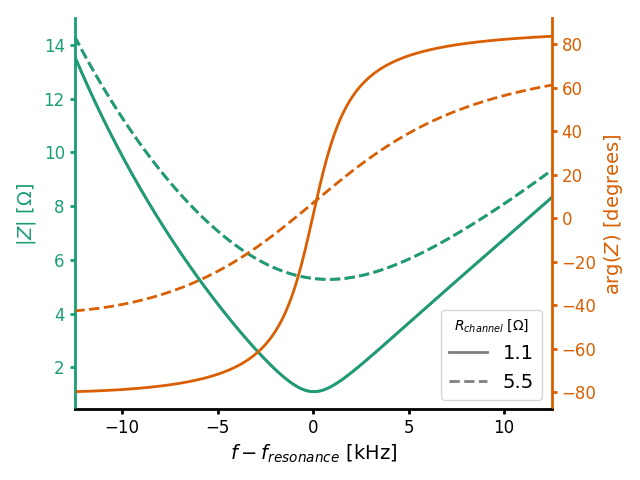}}
  \caption{\textit{Left:} Noise spectral densities for the currents through two channels at 2.8 and 2.9 MHz with differing resistances within the same multiplexer unit.  The measurement was taken at $T$ = 0.3 K and with similar probe currents in both channels.  The lack of asymmetric noise in the more resistive channel shows that the observed noise is modulated by the resonance width.  \textit{Right:} Calculated complex impedances for the two channels measured on the left.  The slopes of the orange curves at zero frequency offset determine the conversion from frequency noise to phase noise for each channel and demonstrate that narrower resonances are more sensitive to frequency noise.}
  \label{fig_frequency_not_phase}
\end{figure}
\begin{figure*}[!t]
\centering
\subfloat{\includegraphics[width=2.3in]{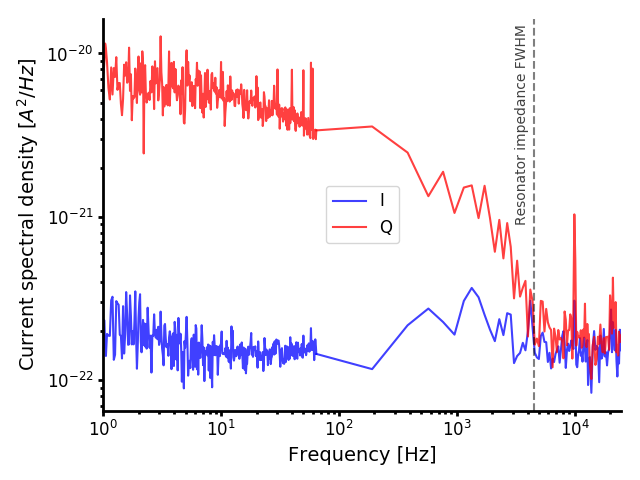}%
\label{tmp1}}
\hfil
\subfloat{\includegraphics[width=2.3in]{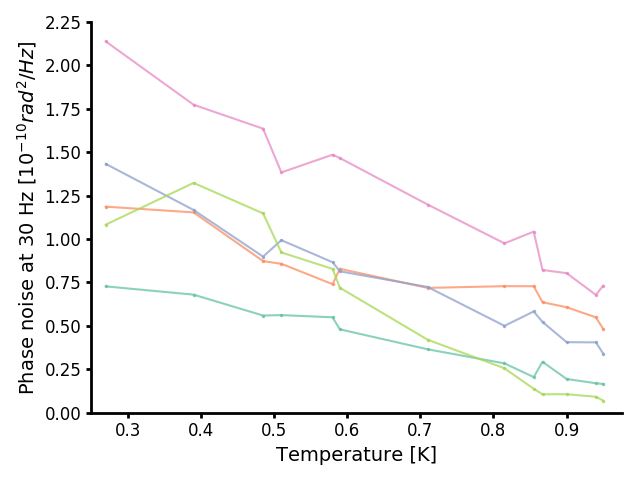}%
  \label{tmp2}}
\hfil
\subfloat{\includegraphics[width=2.3in]{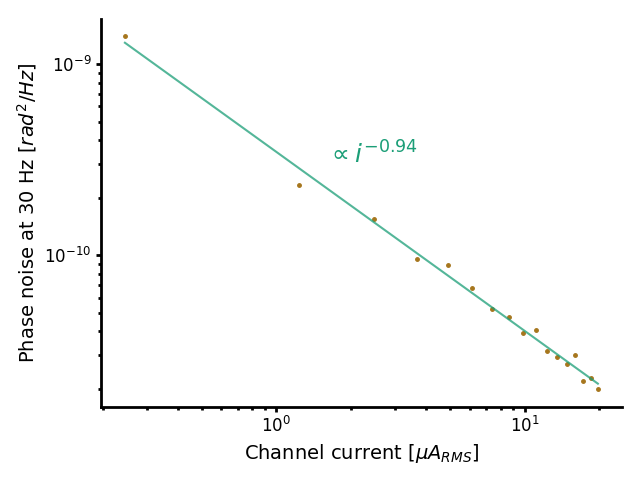}%
  \label{tmp3}}
\caption{\textit{Left:} Spectral density of current through a 2.1 MHz, 1 $\Omega$ channel at $T$ = 0.3 K over a wider frequency range and with the effect of the anti-aliasing filter removed.  The excess asymmetric noise follows a red spectrum out to the resonator bandwidth.  \textit{Center:} Temperature dependence of the phase noise measured with a set of 1 $\Omega$ channels operated at identical probe currents.  Notable channel-to-channel variation is observed, but power law fits yield indices broadly consistent with those reported in the literature for TLS resonator noise.  \textit{Right:} Readout current dependence of the phase noise in a 2.1 MHz, 1 $\Omega$ channel at $T$ = 0.3 K, showing agreement with the $P^{-1/2}=R^{-1/2}I^{-1}\propto I^{-1}$ expectation from TLS fluctuations.}
\label{fig_TLS}
\end{figure*}

The current dependence of the $Q$ noise suggests an underlying phase noise.  In the DfMux readout system, a jitter in the phase $\phi = \tan^{-1}{(Q/I)}$ of the channel current corresponds to a quadrature noise $\delta Q$ via
\begin{equation}
\delta Q = \langle I \rangle \tan{\delta\phi}
\label{eq_phase_to_Q}
\end{equation}
with which we find excellent agreement.\footnote{As later shown in the right panel of Fig. \ref{fig_TLS}, the phase noise $\delta\phi$ itself is a function of $\langle I\rangle$, resulting in a nonlinear dependence of $\delta Q$ on $\langle I\rangle$.}  However, the observed phase noise is in turn due to a frequency noise, as it is observed to be modulated by the conversion factor between frequency noise and phase noise
\begin{equation}
\frac{\delta\phi}{\delta f} = \frac{d(\arg{Z})}{df}\bigg|_{f_{r}}
\label{eq_frequency_to_phase}
\end{equation}
where $f_{r} = (2\pi\sqrt{LC})^{-1}$ is the natural resonance frequency of the channelizing filter and $Z$ is the complex channel impedance.  This is demonstrated in Fig. \ref{fig_frequency_not_phase}, where two channels within the same multiplexer with different sensitivities to frequency jitter can be seen to demonstrate notably different noise asymmetries.

\par
The underlying mechanism behind the observed frequency noise appears to be one common to related systems; as reviewed in \cite{ZmuidzinasReview} and \cite{TLSReview}, superconducting lithographed resonators at $\sim$GHz frequencies generically host fluctuating two-level tunneling states in amorphous dielectrics which couple to the resonator capacitances and induce resonance frequency jitter.  TLS fluctuation noise has a notable set of well-studied dependencies which agree well with our observations.\footnote{TLS populations are also known to modify the resonance frequencies and internal quality factors as functions of temperature and readout power, but these were not convenient to measure with our system.}  We are currently unaware of any other mechanism capable of simultaneously reproducing all the observed noise phenomena.

\par
Importantly, we observe no phase noise correlation between multiplexer channels, which strongly disfavors mechanisms such as temperature or clock drifts and is consistent with the expectation from TLS fluctuators.  Furthermore, TLS noise is known to follow a red spectrum out to the resonator bandwidth where it typically becomes sub-dominant to other noise sources, which we observe as demonstrated in the left panel of Fig. \ref{fig_TLS}.  A characteristic negative temperature dependence of TLS noise is also reported in the literature, typically following a power law of $T^\gamma$ where $\gamma$ varies from $-1.2$ to $-1.7$ \cite{TLSReview}.  Measurements of this dependence in our system were slightly hindered by a lack of direct thermometry and imperfectly constrained thermal gradients as evidenced by the channel-to-channel variation shown in the center panel of Fig. \ref{fig_TLS}, but yield broadly consistent fitted power law indices in the $-0.6$ to $-2.5$ range.  Additionally, the electric field strength in the capacitive part of resonators, or equivalently the average power of the readout tone, is known to suppress TLS noise according to a $P^{-1/2}$ power law.  Indeed this is seen in our resonators, as evidenced by the example in the right panel of Fig. \ref{fig_TLS}.  Finally, TLS population densities are known to be sensitive to fabrication details: material/substrate pairings, residual surface contaminants, oxide layers, etc.  In line with this, we observe on average a factor of $\sim$4 larger phase noise in resonators fabricated from aluminum films with an evaporation process over others fabricated from niobium films with a sputter process.

\section{Impact in the Simons Array}

\par
\begin{figure}[!t]
  \centering
  \includegraphics[width=0.49\linewidth]{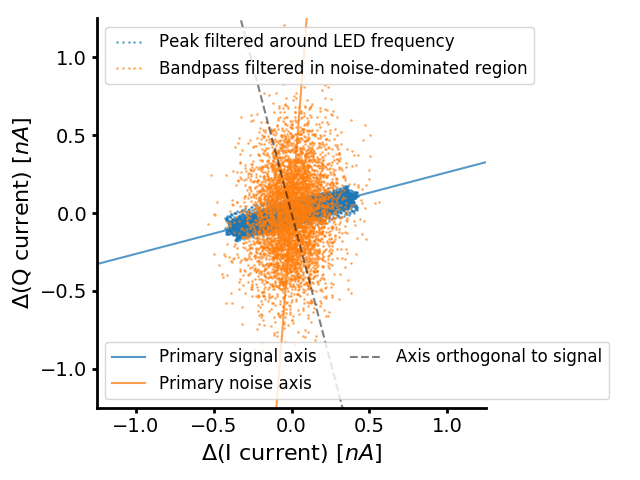}
  \includegraphics[width=0.49\linewidth]{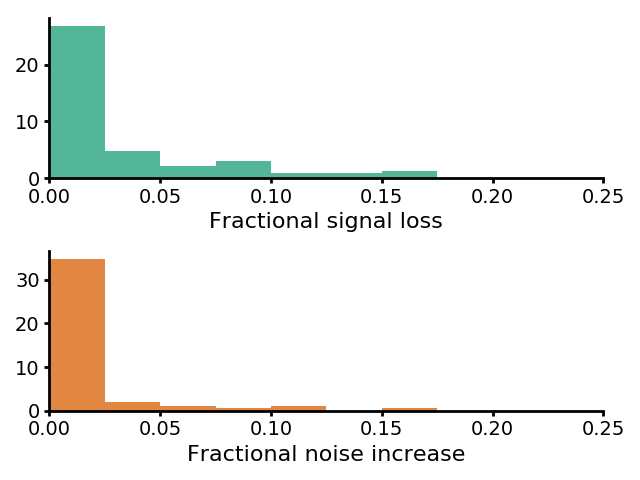}
\caption{\textit{Left:} The complex current measured in a spare Simons Array detector thermally stimulated at 5 Hz, peak-filtered around the stimulation frequency (blue) and bandpassed in a noise-dominated frequency range (orange).  Lines are shown to guide the eye along the principal axis of each data set, showing the non-orthogonality between TLS noise and the detector responsivity.  \textit{Right:} Upper bound of the impact of TLS fluctuations on the Simons Array sensitivity by similarly stimulating 91 such detectors.  The histograms show the fraction of the maximum possible signal lost and the fractional increase over the minimum possible noise as a result of choosing a demodulation basis that optimizes the signal-to-noise ratio.  The relative importance of readout noise in this test is overemphasized due to the lack of photon noise, hence the interpretation of these results as upper bounds.}
\label{fig_SA_impact}
\end{figure}

As is visually apparent in the right panel of Fig. \ref{fig_noise_phenomenology}, a simple change of demodulation basis is sufficient to completely remove the excess TLS noise from the data stream.  However, we observe a complicating effect, namely that the basis which optimizes TLS noise avoidance is in general distinct from the basis optimizing the detector signal.  This phase difference is shown on the left in Fig. \ref{fig_SA_impact} for an outlier channel where the distinction is visually obvious.  Maximizing the detector signal therefore comes with a noise penalty, and vice versa.  Practically, then, the basis which optimizes signal-to-noise should be used.  Presently, the specific mechanism determining the scale of the non-orthogonality of TLS noise and detector responsivity is unclear, but may be a subject of further study.

\par
In laboratory tests of a spare Simons Array TES wafer where detectors were thermally stimulated with out-of-band optical power from an LED in an otherwise dark cryostat, the rotation angles relating the maximum signal-to-noise basis and the default basis where $\langle Q\rangle = 0$ are typically small: centered around zero with a spread of roughly 10 degrees.  To constrain the magnitude of the sensitivity impact on the Simons Array, the signal losses and noise increases relative to a hypothetical system with TLS-free resonators are histogrammed in the right panel of Fig. \ref{fig_SA_impact}, and can be seen to be concentrated at the percent level with a few outliers.  As these test conditions under-represent the contribution from photon noise expected during observations, these results should be treated as upper bounds.  Therefore, while we expect to perform an offline demodulation basis rotation to maximize sensitivity for the Simons Array, we do not expect TLS noise to appreciably contribute to the overall noise level afterwards.

\section{Discussion}

\par
As DfMux is an amplitude readout scheme, it would be insensitive to TLS noise were it not for the observed non-orthogonality between TLS noise and the detector response.  Though the precise mechanism behind the relative angle between signal and noise is unclear, the lack of a fixed relationship is perhaps unsurprising; in other systems susceptible to TLS noise, such as microwave kinetic inductance detectors or microwave SQUID multiplexers, the sensing and readout elements are either co-located in the same structure or coupled via circuitry operating at frequencies significantly below the natural resonance frequency.  In our readout system, the RF probe tone and detector bias are coincident, enabling reactive impedances in the wiring between the detector elements and resonators or in the TES itself to produce phase shifts.

\par
For future systems that may wish to adopt TLS noise mitigation strategies, several options are identifiable.  At a cost of channel density and/or crosstalk performance, the resonator inductance $L$ may be lowered or the TES operating resistance $R_{oper}$ increased, thereby reducing the conversion factor between frequency noise and phase noise given by Eq. (\ref{eq_frequency_to_phase}).  For the practical case in which the detector saturation power is held constant, the phase noise will scale with $L/\sqrt{R_{oper}}$ after accounting for the current dependence of TLS noise.  Reduction of the TLS fluctuations themselves is also an option; a variety of other techniques, including the use of materials that do not form surface oxides, care to remove residual films in the fabrication process, and modification of the capacitor geometries, have been successfully used in other devices and would be applicable for DfMux readout \cite{ZmuidzinasReview}.  Additionally, active feedback on the carrier, similar to that described in \cite{DAN_DEVB} or \cite{FSA}, may also provide a further option for avoiding TLS noise.  As we are unaware of any report of TLS noise in MHz resonators, at the time of design none of these avoidance or mitigation techniques were considered;\footnote{For unrelated reasons, $R_{oper}$ is larger by roughly a factor of 2 and a notable fraction of the resonator capacitors were fabricated with wider finger gaps in the SPT-3G implementation relative to the Simons Array implementation \cite{JoshThesis}.  Both of these likely contribute to the relatively smaller level of observed phase noise in SPT-3G.} we anticipate that implementation will be straightforward.

\par
In summary, we have reported an observation of TLS noise in the DfMux readout system, representing both a previously uncharacterized noise source for the Simons Array and a new regime in which TLS fluctuations have been observed.  We have further shown that the resulting resonator phase noise is not fully orthogonal to the amplitude modulation given by the detector signals.  While we expect to apply an offline phase rotation for mitigation in the Simons Array, the overall sensitivity impact due to TLS noise will be very minor.  Currently, the next generation of CMB cameras are under design, many of which are considering DfMux readout; these and other related TES systems should be aware that TLS noise can manifest in MHz resonators, take care not to exacerbate its effects, and implement mitigation strategies if desired.

\section*{Acknowledgment}
The authors would like to thank Christopher Raum and Benjamin Westbrook for their assistance with the resonator fabrication, as well as the Simons Array and SPT-3G collaborations for their assistance in developing the readout hardware and for providing infrastructure for testing.

\IEEEtriggeratref{9}

%
\bibliographystyle{IEEEtran}


\end{document}